\documentclass[aps,prl,twocolumn,superscriptaddress,footinbib,linenumbers]{revtex4}

\usepackage{graphicx}
\usepackage{dcolumn}
\usepackage{bm}
\usepackage{braket}
\usepackage{exscale}
\usepackage[intlimits]{amsmath}
\usepackage{amsfonts}
\usepackage{amssymb,amscd}
\usepackage{color}
\usepackage{hyperref}
\usepackage{subfigure}
\usepackage[normalem]{ulem}
\usepackage{slashed}
\usepackage{leftidx}
\usepackage{booktabs}

\begin{document}

\title{Scalable Generation of Massive Schrödinger Cat States via Quantum Tunneling}

\author{Han Zhang}
\thanks{These authors contributed equally to this work.}
\affiliation{State Key Laboratory of Quantum Functional Materials, Department of Physics and Guangdong Basic Research Center of Excellence for Quantum Science, Southern University of Science and Technology, Shenzhen 518055, China}
\affiliation{Quantum Science Center of Guangdong-Hong Kong-Macao Greater Bay Area, Shenzhen 518045, China}
\author{Yong-Kui Wang}
\thanks{These authors contributed equally to this work.}
\affiliation{State Key Laboratory of Quantum Functional Materials, Department of Physics and Guangdong Basic Research Center of Excellence for Quantum Science, Southern University of Science and Technology, Shenzhen 518055, China}
\author{Yi Zheng}
\thanks{These authors contributed equally to this work.}
\affiliation{State Key Laboratory of Quantum Functional Materials, Department of Physics and Guangdong Basic Research Center of Excellence for Quantum Science, Southern University of Science and Technology, Shenzhen 518055, China}
\author{Hai-Tao Bai}
\affiliation{State Key Laboratory of Quantum Functional Materials, Department of Physics and Guangdong Basic Research Center of Excellence for Quantum Science, Southern University of Science and Technology, Shenzhen 518055, China}
\author{Bing Yang}
\email[e-mail:]{yangbing@sustech.edu.cn}
\affiliation{State Key Laboratory of Quantum Functional Materials, Department of Physics and Guangdong Basic Research Center of Excellence for Quantum Science, Southern University of Science and Technology, Shenzhen 518055, China}
\affiliation{Quantum Science Center of Guangdong-Hong Kong-Macao Greater Bay Area, Shenzhen 518045, China}


\begin{abstract}
Massive objects in spatial superposition may provide insights into the interplay between quantum mechanics and gravity. Cold atomic interferometers offer a promising platform due to extended matter-wave coherence times and precise controllability. However, high-mass spatial superpositions beyond single atoms have yet to be generated in such setups. Here, we report the scalable realization of high-mass spatial entanglement via quantum tunneling of ultracold atoms in optical lattices. We observe coherent tunneling of bound clusters, forming a composite object with a mass of 608~amu. Full control of the model parameters allows us to mitigate the usual suppression of tunneling with increasing mass. Furthermore, we construct an interferometer to certify the entanglement and use spatially distributed Schr\"{o}dinger cat states to perform quantum-enhanced measurements. These results establish an approach to generating and detecting massive superposition states relevant to studies of quantum gravity.
\end{abstract}
\date{\today}
\maketitle

{\it Introduction.}
Quantum theory and general relativity are the two central frameworks of modern physics, yet they remains fundamentally incompatible.
Macroscopic spatial quantum superpositions have been proposed as potential probes for exploring the interface between quantum mechanics and gravity~\cite{Feynman1957,Bose2025,Marletto2025}.
Recent experiments have achieved high-mass spatial superpositions in platforms including large molecules~\cite{Fein2019} and micromechanical oscillators~\cite{Bild2023}.
However, the spatial extent of these superpositions remains well below nuclear dimensions, rendering their spatial delocalization effectively negligible.
In contrast, cold atom interferometry has demonstrated long-lived coherence in single-atom superpositions, achieving meter-scale spatial separations~\cite{Kovachy2015} and coherence times exceeding one minute~\cite{Panda2024}.
Extending such interferometric techniques to systems with masses beyond individual atoms would enhance their potential for probing quantum aspects of gravity~\cite{Szigeti2021}.

In cold atomic systems, single-atom motional superpositions can be generated via coherent light-atom interactions~\cite{Szigeti2021}.
Entangling multiple atoms requires interatomic interactions, making Bose–Einstein condensates with large atom numbers and extended coherence a suitable platform~\cite{Cirac1998,Andrews1997}.
In this context, quantum tunneling of ultracold atoms offers a pathway to generating spatial entanglement, which is distinct from entanglement in internal spin states~\cite{Jiang2026}.
For weakly interacting atoms, the Josephson effect can yield coherent superpositions of population across spatial modes~\cite{Smerzi1997,Albiez2005}. However, these states generally approximate coherent states and thus offer only a near-classical measurement precision~\cite{Szigeti2021,Pezze2018}.
In strongly interacting systems, correlated tunneling processes involving atom pairs have been experimentally demonstrated~\cite{Folling2007,Preiss2015,Murmann2015,Yang2020a}.
Scaling these approaches to larger atom numbers, however, presents substantial challenges, including the exponential suppression of tunneling strength with increasing mass~\cite{Razavy2013} and enhanced three-body losses in interacting atomic ensembles~\cite{Burt1997}.

Here, we overcome key limitations and experimentally demonstrate a scalable method for generating genuine spatial entanglement among multiple bosonic atoms.
For bound ultracold atoms, we suppress three-body losses and identify a regime in which the tunneling strength remains nearly independent of mass.
Specifically, when the mutual binding energy $U$ (between two atoms) exceeds the single-atom kinetic energy $J_0$ but remains below the barrier height (i.e., $V_0 \gg |U| \gg J_0$), a cluster of $n$ atoms can tunnel through the barrier as a single entity via a high-order perturbative process, with a tunneling strength that scales as $J_0^n/U^{n-1}$.
We find that the dependence remains approximately consistent with the perturbative form even for $J_0/U \sim 1$, where the collective tunneling strength remains finite and experimentally resolvable.

Experimentally, we realize atomic Mott insulators with tunable sub-Poissonian integer filling in optical lattices, where on-site repulsion binds the atoms into an array of clusters.
By precisely stabilizing the superlattice potential, we significantly extend the coherence time of spatially separated entangled states.
Using time-resolved measurements, we directly observe parallel coherent tunneling of hundreds of atomic clusters arranged in a lattice array, each composed of up to 7 atoms (608.4 amu).
The tunneling strength remains nearly constant with increasing mass, offering a promising route toward scaling up to larger masses.
Moreover, the tunneling dynamics generates Schrödinger cat states in position space, which we exploit to achieve enhanced sensitivity to sub-micrometer energy shifts with hertz-level precision, surpassing the standard quantum limit.

\begin{figure}[!tb]
\includegraphics[width=7.9cm]{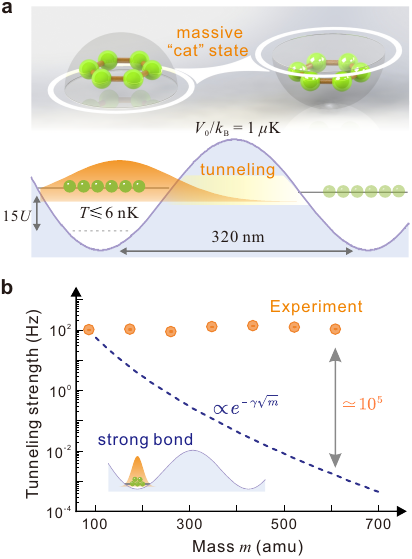}
\caption{Massive Schrödinger cat states induced by quantum tunneling.
(a) Quantum tunneling of ultracold clusters.
In a double-well potential with a barrier height of $V_0/k_B = 1~\mu$K and a width of 320~nm, atoms at temperature below $6$~nK form weakly bonded clusters via on-site interactions.
The matter wave of clusters penetrate into the adjacent well, where the interactions elevate the energy levels, introducing an energy penalty that suppresses competing tunneling channels.
During the tunneling process, genuine entangled states of massive clusters emerge.
(b) Observation of tunneling-strength scaling in massive clusters.
By forming weakly bound clusters of $^{87}$Rb atoms, we observe a clear near-unity scaling of tunneling strength with mass, as indicated by the orange data points.
Each data point represents the mean value $\pm$ SEM, extracted from an average of 216 samples.
In contrast, strongly bonded clusters would feature smaller matter-wave packets, and their tunneling strength is expected to decrease exponentially, as shown by the blue curve.
}
\label{Fig1}
\end{figure}

{\it Theoretical descriptions.}
Quantum tunneling is a fundamental quantum effect arising from the matter-wave properties of objects~\cite{Razavy2013}.
The tunneling of electrons is extensively utilized in devices such as scanning tunneling microscopes, and superconducting qubits for quantum computing.
For a general object with mass $m$ and kinetic energy $K$ traversing a square barrier of height $V_0$ and thickness $d$, the transmission probability is proportional to $\exp(-\gamma \sqrt{m})$, where $\gamma=2\sqrt{2(V_0-K)}d/\hbar$  ($\hbar$ is the reduced Planck constant).
{This relation provides an intuitive picture in which the tunneling strength decays exponentially with increasing mass $m$}, leading to a fundamental question: to what extent can quantum tunneling be realized on macroscopic objects with large masses.

Here, we consider a double-well system as the minimal platform for studying the tunneling effect, as illustrated in Fig.~\ref{Fig1}$\rm{a}$.
For strongly bonded atoms with attractive interaction $|U| \gg V_0$, $U<0$, the atoms form a molecular cluster, leading to a smaller matter-wave packet and an exponential reduction in tunneling strength, scaling as $\exp(-\gamma \sqrt{m})$ (see Fig.~\ref{Fig1}$\rm{b}$ {and Methods}).
In the weakly bound regime ($|U| \ll V_0$), the interaction can be treated as a pseudopotential, allowing the atomic wavefunction to approximate that of a single atom~\cite{Winkler2006,Bloch2008,Lewenstein2012}.
This provides an alternative high-order perturbative path for the tunneling of clusters from the state $\ket{n,0}$ to $\ket{0,n}$ through virtual states $\ket{n-i,i}$, where $i=1,2,\dots,n-1$.
Repulsive interactions ($U > 0$) lift the zero-energy level, thereby energetically detuning intermediate virtual states and effectively suppressing all lower-order tunneling processes.
In the perturbative regime $J_0 \ll U$, the effective tunneling strength of this high-order process is given by $\alpha_n {J_0}^n/U^{n-1}$, where $\alpha_n = n/(n-1)!$ accounts for the boson enhancement (with a factor of $n!$) and energy-shifts of virtual states relative to $\ket{n,0}$ in multiple of $U$.

\begin{figure*}[!tb]
\includegraphics[width=17.4cm]{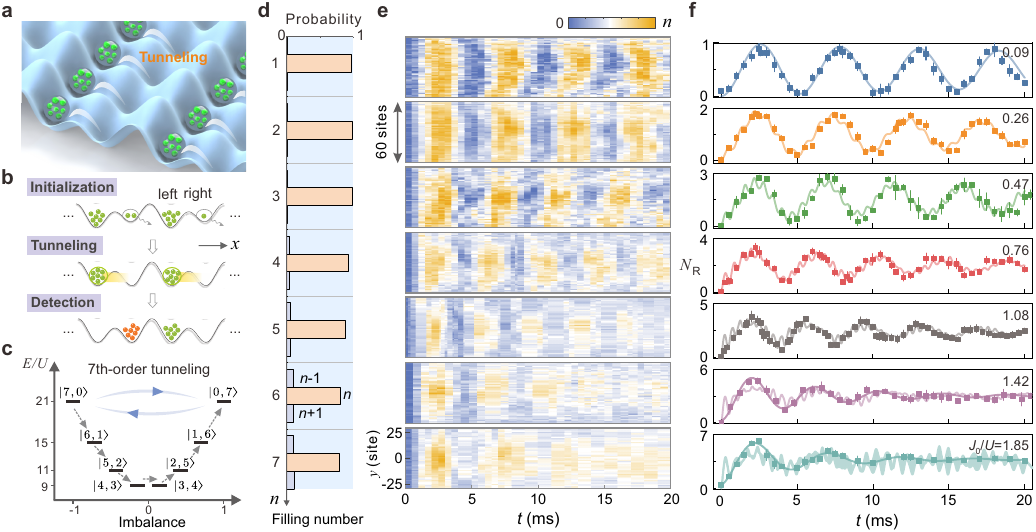}
\caption{Quantum tunneling dynamics.
(a) Experimental platform.
The system consists of array of atomic clusters trapped in an optical superlattice. The atoms are bonded into clusters and tunnel collectively.
(b) Experimental sequence.
The atoms are initialized using a cooling technique, followed by the removal of the right-site atoms. During the evolution stage, the atomic clusters tunnel through the barrier in balanced double-wells. The quantum state is measured by selectively detecting the atoms on the right site, yielding the value $N_R$.
(c) Energy spectrum at $J_0 \ll U$ and high-order tunneling process.
For bonded atoms in the $\ket{7,0}$ state, tunneling to the $\ket{0,7}$ state occurs via a 7th-order process, linked by 6 intermediate virtual states.
(d) Mott insulators with central fillings of 1 to 7.
The probabilities of each component are obtained through parity-projection measurements.
(e) Time-resolved observations of tunneling dynamics.
$In~situ$ measurements are used to detect the time evolution and spatial distribution of the atomic clouds.
The signal is averaged over 14 sites along the $x$-axis.
(f) Coherent tunneling of atomic clusters.
Data points represent the spatial average of $N_R$ over the central 14 sites in (e). The Hubbard parameter $J_0/U$ are labeled in the right corner of each plot.
Data points represent mean values $\pm$ SD, alongside theoretical predictions (solid lines). The data are obtained from $1.4\times10^3$ samples.
Fitting curves for the six- and seven-atom clusters are also included.
The theoretical calculations incorporate phenomenological decay rates and occupation imperfections.
For the seven-atom cluster, $U$ is taken from experimental calibration.
}
\label{Fig2}
\end{figure*}

We create double-well potentials and realize weakly bonded clusters using an optical superlattice.
Ultracold bosons in the optical lattices can be described by the ground-band Bose-Hubbard model~\cite{Jaksch1998},
\begin{equation}
\label{eq:BHM}
\hat{H}_0 = \sum_{ j } \left[-J_0\left(\hat{a}^\dagger_{j} \hat{a}_{j+1} +\text{h.c.}\right)+ \frac{U}{2} \hat{n}_{j} (\hat{n}_{j} - 1) +\varepsilon_j \hat{n}_{j}\right],
\end{equation}
where $\hat{a}^\dagger_{j}~(\hat{a}_{j})$ denotes creation (annihilation) operator at site $j$, with $\hat{n}_{j}=\hat{a}^\dagger_{j} \hat{a}_{j}$ as the atom number operator.
The term $\varepsilon_{j}$ represents the on-site energy and chemical potential.
In this system, two-body repulsive on-site interactions lead to the binding energy term~\cite{Winkler2006}.
Under this model, a superfluid-to-Mott-insulator phase transition can yield a uniform region with an integer atom number per site~\cite{Jaksch1998,Greiner2002}, providing an ideal starting point for studying quantum tunneling.

By adjusting the double-well superlattice potentials ~\cite{SebbyStrabley2006,Folling2007,Yang2020}, we reach the regime $V_0 \gg U > J_0$.
As illustrated in Fig.~\ref{Fig1}$\rm{a}$, we set the barrier height to $V_0/k_B \simeq 1\ \mu \text{K}$, the binding energy to $U/k_B \simeq 40 ~\text{nK}$, and a tunable bare tunneling strength $J_0$, achieving the weakly bonded condition.
To characterize the collective tunneling dynamics of bonded atoms initialized in the state $\ket{n,0}$, we can employ a simplified two-mode model to describe the perturbative regime,
\begin{equation}
\label{eq:Hamiltonian}
\hat{H}_{\text{c}} = -J_{n}  \left[\big(\hat{a}_{L}^\dagger\big)^n\big(\hat{a}_{R}\big)^n + \text{h.c.} \right],
\end{equation}
where the notation $\text{L}/\text{R}$ denote the corresponding operators acting on the atoms at the left or right sites, respectively, and the coupling strength is given by $J_n={J_0}^n/\left[(n-1)!^2 U^{n-1}\right]$.
We experimentally measure the tunneling strength as a function of mass, {with results in good agreement with the theoretical model (see Fig.~\ref{Fig1}\rm{b}), where the coupling parameters $J_0$ and $U$ are slightly adjusted}.
In the crossover regime $J_0 \sim U$, the perturbative scaling provides intuition, but the full dynamics are solved exactly using the Bose-Hubbard model. The gap of $(n-1)U$ between the initial state and the first intermediate state suppresses population of intermediate configurations, thereby protecting collective tunneling as the dominant process, while the quantitative scaling prefactors $\alpha_n$ are moderately renormalized.

{\it Experiment.}
The experiment starts with a nearly pure Bose-Einstein condensate of ${}^\text{87}\text{Rb}$ atoms, containing $\sim 2 \times 10^5$ atoms in the $\ket{F=1,m_F=-1}$ state.
The atoms are loaded into a single layer of pancake-shaped standing waves along the $z$-axis, with the atom number in this quasi-two-dimensional (2D) system adjustable via further evaporation (see Supplementary Information).
This 2D gas is then adiabatically loaded into a square optical lattice in the $x-y$ plane.
Along the $y$-axis, the lattice is created using counter-propagating lasers with a wavelength of $\lambda_s = 767$ nm, while along the $x$-axis, a superlattice composed of lasers with wavelengths of $\lambda_s$ (short-lattice) and $2\lambda_s$ (long-lattice) divides the trapped atoms into a series of double wells.
The intensity maxima of these two lattice patterns are aligned, creating a balanced double-well superlattice configuration, as shown in Fig.~\ref{Fig2}\rm{a}.
The barrier height $V_0$ can be finely tuned by adjusting the lattice depths, and the barrier width is slightly less than $\lambda_s/2$, reaching 320 nm.
The superlattice is precisely engineered to allow full control over the parameters in the model described by Eq.~\ref{eq:BHM}.


\begin{figure*}[!tb]
\includegraphics[width=16.3cm]{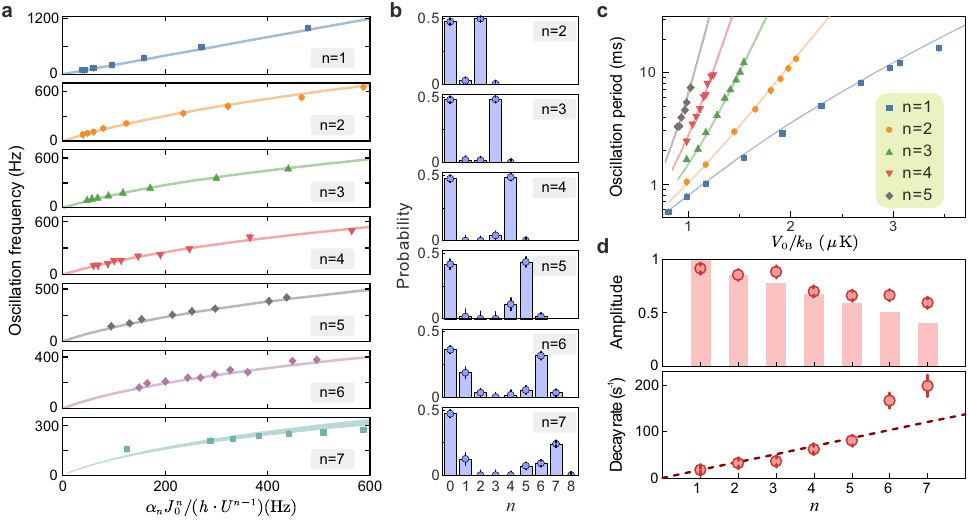}
\caption{Scaling of tunneling strength.
(a) Oscillation frequencies of tunneling dynamics.
The frequencies of tunneling atomic clusters are extracted from the data, exemplified in Fig.~\ref{Fig2}\rm{f}.
These frequencies are presented as a function of the Hubbard parameters in the form of $\alpha_n J_0^n/U^{n-1}$.
Solid lines represent theoretical calculations based on calibrated Hubbard parameters.
The data are obtained from $8.5\times 10^3$ samples.
(b) Collective tunneling and full counting statistics.
We measure the atom-number distributions of the left site at the first quarter of the tunneling period. The superposition states with cluster size from $n=2$ to $n=7$ are characterized. The measured probabilities exhibit a bimodal structure with dominant peaks at $p_0$ and $p_n$. The data are obtained from $1.5\times10^3$ samples.
(c) Oscillation periods and barrier height.
A logarithmic plot is used to illustrate the exponential dependence of the period on the barrier height.
Further analysis reveals a linear dependence of the exponent on cluster size $n$.
The data are obtained from $7.3\times10^3$ samples, and solid lines represent theoretical results.
(d) Normalized tunneling amplitude and coherence time.
The tunneling oscillation amplitudes, normalized by $n$, decrease slowly with increasing mass.
The histograms represent theoretical calculations.
The decay rate increases with $n$, and the dashed line illustrates a linear trend.
For $n=1$, the measured coherence time is $\tau^{(1)} = 58(8)$ ms.
The data are obtained from $1.4\times10^3$ samples.
All data points in this figure are shown as mean values $\pm$ SEM.
}
\label{Fig3}
\end{figure*}


We initialize the system by utilizing the integer filling property of the Mott insulator and a site-resolved atom manipulation technique.
In this superlattice, we implement a cooling method for the Mott insulators by transferring entropy to nearby superfluid reservoirs~\cite{Yang2020}.
The central filling factor $n$ can be tuned from 1 to 7, with the corresponding temperatures measured between $2.2(2)$~nK and $6.2(1)$~nK.
The repulsive on-site interaction is around $U/h= 835(6)$~Hz, binding the particles together during the collective tunneling.
Three-body loss is relatively low, with the $1/e$ lifetime of the sample at $n=7$ reaching 1.8(1)~s.
Such a low rate of three-body loss may primarily result from suppression effects in the strongly interacting regime~\cite{Tolra2004}.
To prepare the initial state with an array of atomic clusters occupying only the left sites, we selectively flip the internal states of atoms on the right sites and subsequently remove them using resonant light, as illustrated in Fig.~\ref{Fig2}\rm{b}.

We monitor the tunneling dynamics of bonded clusters in an array of isolated double wells.
In the insulating state, tunneling is suppressed by setting high barriers in all directions.
Along the $x$-direction, the long-lattice depth is kept at $8.87(4)E_r$ to prevent inter-well tunneling between adjacent double wells.
Here, $E_r = h^2/(2 m_0 \lambda_s^2)$ denotes the recoil energy, and $m_0$ is the atomic mass of ${}^{87}$Rb.
For double-well systems prepared in the $\ket{n,0}$ state, we lower the barrier and enable atoms to tunnel through.
After a set interval, we rapidly raise the barrier to freeze the tunneling dynamics and prepare for detection.
Figure~\ref{Fig2}\rm{c} shows the energy spectrum of a seven-atom bonded cluster in the Fock state basis, where tunneling is effectively a 7th-order perturbation process.
In the balanced double well, the states $\ket{7,0}$ and $\ket{0,7}$ are degenerate, while each intermediate state has a lower energy, acting as virtual states that do not noticeably populate during the evolution.
Since maintaining the balanced configuration is critical to keeping the $\ket{7,0}$ and $\ket{0,7}$ states resonant, we stabilize the energy bias $\Delta \equiv \varepsilon_{\text{R}} - \varepsilon_{\text{L}}$ between wells to $2.5(1) \times 10^{-4}$ level relative to $V_0$.

We measure quantum states by selectively detecting the atom number on the left or right sites using $in~situ$ absorption imaging.
The number distributions of the initial states are shown in Fig.~\ref{Fig2}\rm{d}, revealing the characteristic sub-Poissonian statistics of atomic Mott insulators.
After the tunneling dynamics, we flip the right-site atoms to the $\ket{F=2,m_F=-2}$ state and then detect the quantum mechanical expectation values $N_{\text{R}}$.
Fig.~\ref{Fig2}\rm{e} presents the time-resolved measurements, where the tunneling dynamics exhibits pronounced coherent oscillations over 20~ms.
A spatially inhomogeneous distribution across 60 lattice sites is observed, mainly due to the Gaussian intensity profile of the lattice beams and the non-uniform atomic density distribution.


\begin{figure*}
\includegraphics[width=15.7cm]{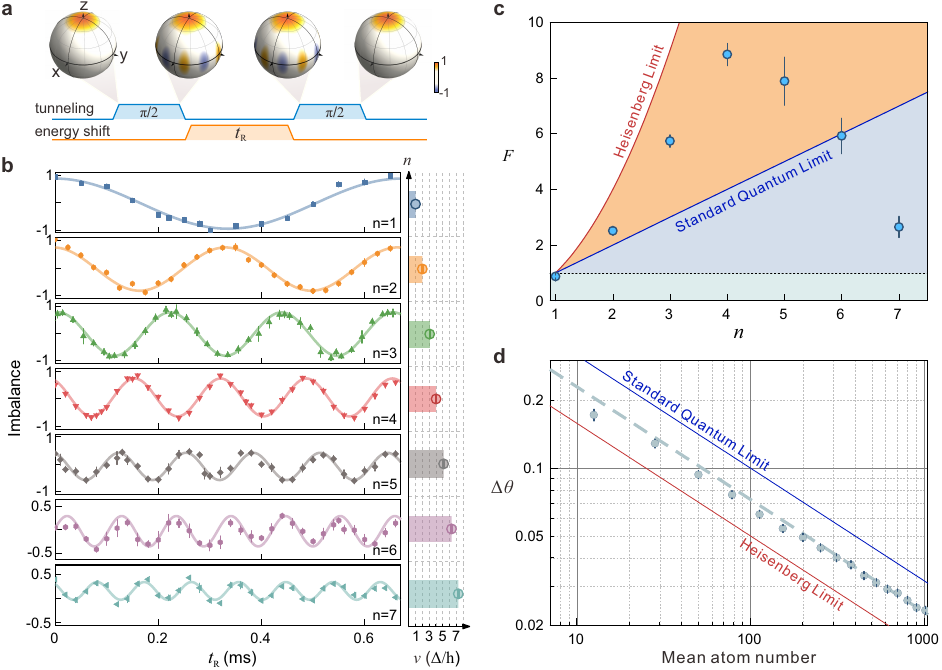}
\caption{Interferometry with Schr\"{o}dinger cat states.
(a) Ramsey interferometer.
We represent the quantum states on the Bloch sphere with the Wigner distribution.
Tunneling of clusters generates entangled states, exemplified by the ideal NOON state for $n=5$.
This state evolves under a site-dependent energy shift, which rotates its phase by $\theta=\pi/5$ (as an example).
In the detection stage, a second tunneling event acts as an equivalent $\pi/2$-pulse in this Ramsey interferometer.
(b) Detecting an energy shift with superposition states.
The atom number imbalance between the left and right sites, expressed as $\left(N_{\text{R}}-N_{\text{L}}\right)/\left(N_{\text{R}}+N_{\text{L}}\right)$, is used to track the evolution.
Each data point represents at least 5 samples.
Solid lines represent sinusoidal fits, and the oscillation frequencies are given in units of $\Delta/h$, as shown in the right panel. The mean values of imbalance are slightly shifted from 0 due to other filling components in the initial state.
(c) {Quantum} Fisher information.
We extract the oscillation amplitude from the measurements shown in (b) and calculate the Fisher information for each state based on the occupation probabilities.
The Heisenberg limit scales as $F=n^2$ (red line), the standard quantum limit is $F=n$ (blue line), and the dashed line denotes the single-atom detection limit.
Superpositions of up to 5 atoms lie above the standard quantum limit.
The data are obtained from $1.5\times10^3$ samples.
(d) Precision measurements with the $n=4$ NOON state.
The data are obtained from $225$ samples, where atom numbers are counted within progressively larger regions to incorporate more double-well units for statistical analysis.
The results show power-law dependence (dashed blue line) and enhanced sensitivity with increasing entangled unit number.
All data points in this figure are shown as mean values $\pm$ SEM.
}\label{Fig4}
\end{figure*}


To quantify the tunneling dynamics, we focus on the central region of the system, where the interaction strengths and filling factors are uniform.
From the measurements in Fig.~\ref{Fig2}\rm{f}, we extract the dominant oscillation frequency, amplitude, and coherence time, as exemplified by the fitted curves for six-atom and seven-atom tunneling.
We adjust $J_0/U$ to make the oscillation frequencies of different states comparable.
In the single-atom case, the tunneling oscillation exhibits a full-amplitude sinusoidal pattern, with a coherence time of 58(8)~ms, among the longest reported for quantum tunneling~\cite{Yang2020,Bojovic2025}.
For $n$-atom clusters, oscillation amplitudes approaching $n$ suggest that the dynamics arise from collective multi-particle tunneling.
The small-amplitude, high-frequency oscillations superimposed on the primary tunneling dynamics are attributed to the incomplete suppression of unwanted tunneling processes.
We theoretically calculate the dynamics using exact diagonalization methods that incorporate extended Hubbard terms~\cite{Dutta2015}.
The experimental results show excellent agreement with the theoretical predictions.

We extract the effective tunneling stength from the oscillation frequencies.
According to perturbation theory, when $J_0/U \ll 1$, the tunneling strength is $ \alpha_n {J_0}^n / U^{n-1}$.
In the crossover regime $J_0/U \sim 1$, perturbation theory is no longer strictly limited to the $n$-th order process.
Nevertheless, energy degeneracy between $\ket{n,0}$ and $\ket{0,n}$, together with the finite gap to other intermediate states, suppresses leakage and allows collective tunneling to remain the dominant process.
We theoretically calculate the frequencies with full Bose–Hubbard model and plot the results in Fig.~\ref{Fig3}\rm{a}.
In a wide range of coupling strengths, the oscillation frequency exhibits an almost linear dependence on ${J_0}^n / U^{n-1}$, with an $n$-dependent prefactor.
Experimental data agree with theoretical predictions,
supporting this characteristic scaling behavior for atomic clusters.

We perform full counting statistics measurements of the quantum state during the tunneling dynamics. After the first quarter of each tunneling period, where the coherent population transfer is most pronounced, the state is projected onto the Fock basis. Exploiting the tunability of the superlattice, we can resolve the occupation probabilities in each well, thereby obtaining the full counting statistics~\cite{Cheinet2008,Kwan2024} (see Supplementary Information for details).
As shown in Fig.~\ref {Fig3}\rm{b}, the number distribution on the left well exhibits a pronounced bimodal structure, with dominant peaks at $p_0$ and $p_n$ and only weak residual population in intermediate sectors. The remaining leakage at higher atom numbers is mainly attributed to imperfections in the initial state preparation and subsequent manipulations. This behavior indicates that the dynamics are largely confined to the $\ket{n,0}$ and $\ket{0,n}$ manifold, supporting the interpretation of collective tunneling.

While the mass-dependent scaling is tunable, the bare tunneling rate $J_0$ still follows an exponential decay with increasing $V_0$ (since $\gamma \propto \sqrt{V_0}$ by neglecting the kinetic energy of cold atoms).
This exponential behavior is evident for clusters of varying mass, as shown in Fig.~\ref{Fig3}\rm{c}. When atoms form clusters, the exponent is further amplified by a factor of $n$, as reflected in the steeper slope of the tunneling period.
For heavier clusters, the experimental regime becomes more constrained as $V_0$ rises, further corroborating the collective tunneling behavior of atoms.

The mass dependence of the tunneling strength is clearly evidenced in Fig.~\ref{Fig1}\rm{b}, where the measured values follow a unity scaling behavior.
In addition to the leading-order dependence ${J_0}^n / U^{n-1}$, the reduction of tunneling strength due to prefactor $\alpha_n = n/(n-1)!$ can be facilitated by setting $J_0/U \gtrsim 1$.
It is important to note that the energy gap {suppressing unwanted single-atom tunneling is set by the separation between the initial state $\ket{n,0}$ and the first intermediate state $\ket{n-1,1}$, which equals $(n-1)U$ rather than $U$.}
This distinction highlights that our strongly interacting system operates beyond the mean-field Josephson regime~\cite{Albiez2005,Pezze2018}, where the energy gap is fixed at $U$, thereby preventing access to the parameter regime required for achieving near-unity scaling with mass.

The oscillation amplitude and coherence time provide critical insights into the spatially entangled states generated during the tunneling dynamics.
As shown in Fig.~\ref{Fig3}\rm{d}, the normalized oscillation amplitude decays much more slowly than the conventional exponential form, indicating efficient entanglement preparation using our protocol.
Theoretical calculations suggest that even at a total mass of $10^4$ amu, the oscillation amplitude remains clearly resolvable (see Supplementary Information).
The observed decay rate scales approximately linearly with
$n$, implying that decoherence primarily arises from cumulative single-particle contributions.
By technically extending the coherence time~\cite{Panda2024}, our system demonstrates promising prospects for scaling to larger atomic clusters.

{\it Precision measurements.}
For a spatial superposition state expressed as $\left(\ket{n,0} + \ket{0,n}\right)/\sqrt{2}$, this corresponds to a maximally entangled Schr\"{o}dinger cat state, commonly referred to as a NOON state~\cite{Lee2002}.
The spatial separation of massive atoms in $\ket{n,0}$ and $\ket{0,n}$ makes such entangled states particularly sensitive to spatial energy shifts, while their quantum correlations further enhance the phase sensitivity $\Delta\theta$~\cite{Braunstein1994,Pezze2018}.
For an ideal NOON state, the phase sensitivity could reach the Heisenberg limit, $\Delta\theta=1/n$, surpassing the standard quantum limit, $\Delta\theta=1/\sqrt{n}$.
Compared to NOON states in photonic systems~\cite{Thomas2022} or spatially non-separated qubits~\cite{Jiang2026}, spatially distributed massive entanglement offers the unique potential to probe mass-sensitive fields and atom-sensitive interactions.

Spatial superposition states emerge during the coherent tunneling dynamics. At a quarter oscillation period, the generated state is  a near-NOON state, whose ideal form is  $\left(\ket{n,0} + \text{i}\ket{0,n}\right)/\sqrt{2}$.
We construct a Ramsey interferometer and exploit the multiparticle entanglement to probe a sub-micron energy shift generated by the optical superlattice~\cite{Yang2017}.
As shown in Fig.~\ref{Fig4}\rm{a}, tunneling is harnessed to generate entanglement and detect the accumulated phase, analogous to the $\pi/2$-pulse in the Ramsey interferometer protocol.
The energy shift $\Delta$ between the left and right sites induces a single-atom phase accumulation over time $t_\text{R}$, described by $\theta = t_\text{R} \Delta/ \hbar$.
This evolution imprints an $n-$fold enhanced relative phase on the entangled state, which can be expressed as $\left[\ket{n,0} + e^{\mathrm{i}(n \theta + \pi/2)}\ket{0,n}\right]/\sqrt{2}$.
The entangled state can be visualized on the Bloch sphere, where a rotation of $2\pi/n$ restores the state, highlighting the enhanced phase sensitivity of $n$-atom entanglement.

Fig.~\ref{Fig4}$\rm{b}$ shows the detection results obtained using quantum superposition states.
The single-particle oscillation frequency is measured to be $\Delta/h = 1.49(2)$~kHz, corresponding to the energy shift between adjacent wells.
For clusters containing $n$ atoms, the oscillation frequency scales linearly with $n$, providing additional evidence for the formation of $n$-atom NOON states via tunneling.
{Due to imperfections in state preparation, the achieved fidelity is reduced from the ideal case.}
We utilize Fisher information $F$ to assess the measurement precision, as shown in Fig.~\ref{Fig4}\rm{c}, where the four-atom cluster exhibits the most significant enhancement.
Here, the Fisher information and phase sensitivity exceed the standard quantum limit by 3.4(2) dB and 1.7(1) dB, respectively.
Taking advantage of hundreds of entangled-state arrays generated in parallel during the tunneling process, we further enhance the sensitivity of spatially resolved measurements.
By employing 254 copies of the four-body NOON state, we achieve a phase sensitivity of 0.023(1), corresponding to an energy shift of $h \times 1.4(1)$~Hz over a coherence time of 16(1)~ms, enabling precise detection of sub-micrometer spatial features (see Fig.~\ref{Fig4}\rm{d}).

{\it Conclusion and Outlook.}
Our experiment demonstrates that bonded atoms with a total mass of 608.4 amu can undergo quantum tunneling as a single entity.
To the best of our knowledge, within the category of center-of-mass tunneling of indivisible objects, this corresponds to the largest mass reported so far.
Unlike the well-established exponential decay of tunneling strength with increasing mass, we have developed a method to scale up the mass and observed a near-unity scaling behavior, indicating the feasibility of extending to larger clusters.
Implementing the optical lattice inside an optical cavity could improve the coherence time by up to three orders of magnitude~\cite{Panda2024}.
This approach could extend the tunneling mass by approximately two orders of magnitude, reaching total masses on the order of $10^4$ amu.
Multichannel tunneling~\cite{Meinert2014,Yang2020a} and the coherent connection of entangled states may enable further scaling to larger atomic ensembles.

This achievement contributes to our understanding of quantum tunneling in large-mass objects and open a pathway to addressing the fundamental question of how large a mass can exhibit tunneling effects.
The resulting quantum superposition states offer valuable applications for quantum metrology~\cite{Pezze2018}, particularly in measuring mass-sensitive fields such as gravitational field.
The massive entangled clusters with large spatial separations~\cite{Kaufman2014,Murmann2015}, {when further scaled up toward the Bose-Einstein condensate regime}, could be employed to test fundamental wave-function collapse models~\cite{Bassi2013}.
Furthermore, the high-order interactions and high-fidelity entangling operations in this strongly correlated system may give rise to unconventional quantum phases~\cite{Bloch2008,Lewenstein2012,Dutta2015}, enabling advanced applications in quantum simulation~\cite{Daley2022,Halimeh2025} and quantum information processing.

{\textbf{Acknowledgments:}}
We thank J. Halimeh, P. Hauke, S. Jochim, X. J. Liu, J.-W. Pan, B. Wu, H.-Z. Zhao, Q. Zhou, P. Zoller, and our colleagues at SUSTech for insightful discussions.
This work is supported by the National Key R\&D Program of China (Grant No. 2022YFA1405800), NSFC (Grant No. 12274199), Shenzhen Science and Technology Program (Grant No. KQTD20240729102026004), Guangdong Major Project of  Basic and Applied Basic Research (Grant No. 2023B0303000011) and Guangdong Provincial Quantum Science Strategic Initiative (Grant No. GDZX2204003, GDZX2303002, GDZX2304006, GDZX2405006).

{\textbf{Author contributions:}}
B. Y. conceived, designed, and supervised the experiment.
B. Y., Y.-K. W., H.Z., Y.Z., and H.T. B. built up the apparatus.
H. Z., Y. Z., B. Y. and Y.-K. W. performed the experiments and analyzed the data.
B. Y. and H. Z. wrote the manuscript, with input from all authors.

{\textbf{Competing interests:}}
The authors declare no competing interests.


\vspace*{0.5cm}


\clearpage
\onecolumngrid
\vspace*{0.5cm}
\begin{center}
    \textbf{SUPPLEMENTARY INFORMATION}
\end{center}
\vspace*{0.5cm}
\twocolumngrid
\setcounter{equation}{0}
\setcounter{figure}{0}
\makeatletter
\makeatother
\renewcommand{\theequation}{S\arabic{equation}}
\renewcommand{\thefigure}{S\arabic{figure}}
\renewcommand*{\theHequation}{\theequation}
\renewcommand*{\theHfigure}{\thefigure}

\section{Generation of ultracold atomic clusters}

\begin{figure}[!tb]
\includegraphics[width=8cm]{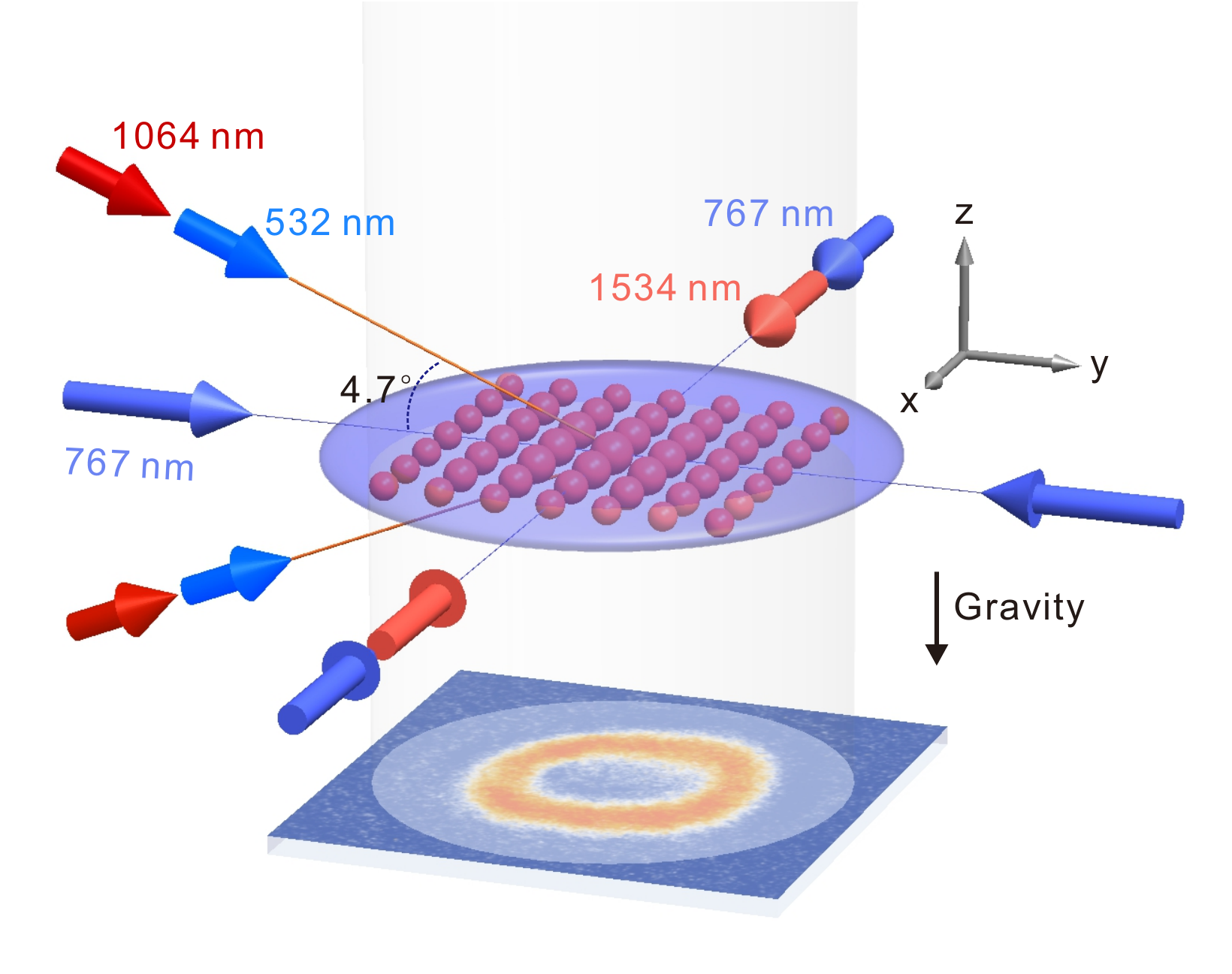}
\caption{Sketch of the experimental setup.
A single layer of a 2D system is created by loading ultracold atoms into a pancake-shaped lattice formed by the interference of 532 nm and 1064 nm laser beams. Retro-reflected laser beams in the $x$ and $y$ directions create the corresponding optical lattices. A blue-detuned lattice with a wavelength of 767 nm is applied in both the $x$ and $y$ directions, while a red-detuned lattice with a wavelength of 1534 nm is used to create a double-well superlattice along the $x$-direction. This setup enables the superfluid-to-Mott insulator transition in the 2D atomic gas. Atomic clouds are imaged via absorption imaging along the $z$-axis, which is aligned with the gravity direction.
}\label{FigS1}
\end{figure}

We generate ultracold atomic clusters by cooling $^{87}$Rb atoms to the quantum degenerate regime, followed by binding the atoms in optical lattices.
Using laser cooling and evaporative cooling techniques, we achieve a nearly pure Bose-Einstein condensate with $\sim2\times10^5$ atoms in the $5S_{1/2}\ket{F=1, m_F=-1}$ state.
The condensate is then compressed along the $z$-direction with an elliptical laser beam, resulting in a Thomas-Fermi radius of $\sim 2~\mu$m.
The atoms are adiabatically loaded into a single layer of a pancake-shaped optical superlattice, which is formed by the overlap of two interference patterns from 1064 nm and 532 nm lasers, intersecting at an angle of $9.3^\circ$.
As shown in Fig.~\ref{FigS1}, the standing wave period along the $z$-axis for the 532 nm interference is 3.3~$\mu$m, while the 1064 nm lattice has a period twice as large.
The superlattice is carefully designed and controlled to ensure the atoms are loaded into a single two-dimensional (2D) system.
The superlattice trap frequency along the $z$-axis is $2\pi \times$ 5.65(4) kHz, ensuring that the system enters the 2D regime.
Due to some heating during compression, we perform additional evaporative cooling to reduce the temperature in this 2D system.

To enter the strongly interacting regime, we load the atoms into optical lattices along the $x$- and $y$-axes (see Fig.~\ref{FigS1}).
The optical lattice along the $x$-axis is a superlattice, with a potential given by $V(x) = V_s \cos^2(2\pi x/\lambda_s) - V_l \cos^2(2\pi x/\lambda_l + \varphi)$, where $\lambda_s = 767$ nm and $\lambda_l = 1534$ nm are the wavelengths of the short- and long-lattices, and $V_s$ and $V_l$ are their respective depths.
The phase $\varphi$ determines the superlattice structure.
To ensure precise phase control, the short-lattice laser is generated by up-converting the long-lattice laser, where the relative frequencies of the two lasers are stabilized to within a Hertz-level precision.
The phase $\varphi$ is finely tuned by adjusting the relative frequency between the two lasers using acousto-optical modulators.
Additionally, the 767 nm laser has a linewidth below 10 kHz.
Compared to previous work in Ref.~\cite{Yang2020}, the frequency fluctuations between the superlattice lasers have been essentially eliminated, with a suppression of several orders of magnitude.

\begin{figure*}[htbp]
\centering
\includegraphics[width=14cm]{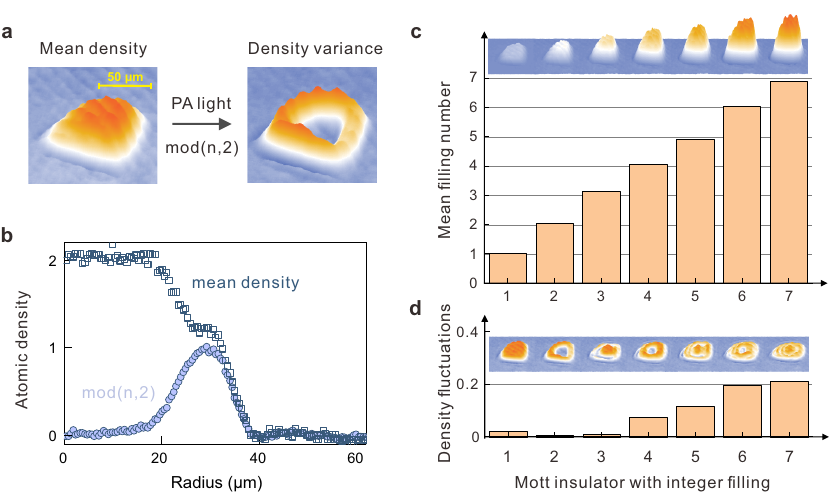}
\caption{Generation of atomic Mott insulators.
(a) $In~situ$ distribution of a Mott insulator. The density distribution of a Mott insulator with central filling $\bar{n} = 2$ in a harmonic trap is shown. The site occupancy is projected onto its parity (odd or even) using photoassociation (PA) excitation.
(b) Density profiles. From the atomic density shown in Fig.~\ref{Fig2}a, a line-cut region is selected for detailed analysis, carefully chosen to avoid distortions due to imperfections in the trapping potential.
The density profiles clearly exhibit the characteristic shell structure.
(c) Mean density. The average filling at the center of the harmonic trap is measured for Mott insulators with filling factors ranging from $n = 1$ to $n = 7$.
(d) Density fluctuations. Fluctuations of the site occupation, obtained via the parity-projection method, are presented.
These measurements enable the extraction of the atomic temperature.
}\label{FigSMI}
\end{figure*}

We utilize the integer filling property of the Mott insulator states to deterministically prepare $n$-atom bonded clusters.
The Mott insulator phase is realized by driving a quantum phase transition governed by the Hubbard model in the 2D system.
To improve the filling factor of the Mott insulator, we apply additional cooling in the optical lattices, as demonstrated in Ref.~\cite{Yang2020}.
Following the cooling sequence, a parity-projection measurement is employed to assess the temperature and filling factor of the Mott insulator states, as shown in Fig.~\ref{FigSMI}\rm{a}.
This technique involves a photoassociation process to excite atomic pairs into unstable molecular states.
We use a laser red-detuned 13.6~cm$^{-1}$ from the D2 line, frequency-locked via a transfer cavity, to drive the transition of atoms to the $v=17$ vibrational state in the $0_g^{-}$ channel.
After applying the laser light for 20~ms, the atomic occupation number $n$ is projected onto its parity, $\mathrm{mod}(n,2)$, distinguishing between odd and even fillings.

\vspace*{0.5cm}

We detect the \textit{in situ} atomic density using highly saturated absorption imaging.
In the Hubbard regime, where lattice sites are occupied by multiple atoms, resolving the two-dimensional atomic distribution requires fast imaging with high saturation.
We precisely calibrate the saturation effects and apply a modified Beer–Lambert law to extract accurate atomic densities. Additionally, Mott insulator states with well-defined integer filling serve as benchmarks for validating our imaging calibration.
Under the local density approximation, the temperature of the atomic ensemble can be inferred from the \textit{in situ} density profile, as illustrated in Fig.~\ref{FigSMI}\rm{b}.
The flat-top and step-like features of the Mott insulator indicate the low-temperature nature of the system.

To cool the central region to the Mott state with filling $n\geq2$, we use a superfluid reservoir with a mean density of $\bar{n}=1.5$.
This configuration blurs the shell structures for higher fillings $n\geq3$, making it difficult to extract temperature from the global density distribution.
To mitigate this issue, we restrict our analysis to the central, uniform region of the atomic cloud, where the local density remains approximately constant.
In this region, we extract the temperature and chemical potential by comparing the measured mean density and density fluctuations with predictions from Bose statistics.
As shown in Fig.\ref{FigSMI}\rm{c}, Mott plateaus with integer fillings from  $n=1$ to $n=7$ are clearly resolved.
The corresponding density fluctuations, obtained after applying the parity projection, are shown in Fig.\ref{FigSMI}\rm{d}. These results allow us to reconstruct the number statistics of the Mott insulator, as presented in Fig.~\ref{Fig2}\rm{d}.

\begin{figure*}[!htbp]
\centering
\includegraphics[width=12cm]{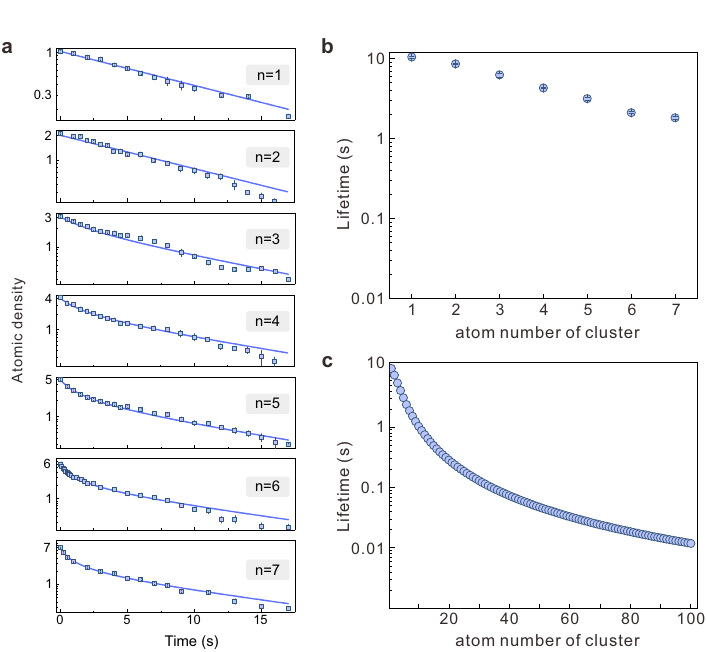}
\caption{Lifetimes of atomic clusters.
(a) Lifetime measurements of $n$-atom clusters.
Starting from Mott insulators with near-uniform filling, we monitor atom number decay in deep optical lattices.
For single and double occupancy, the decay follows an exponential trend consistent with one-body losses.
For clusters with $n \geq 3$, rapid initial decay indicates the contribution of three-body losses.
Solid lines represent fits incorporating both one-body and three-body decay processes.
All data points in this figure are shown as mean values $\pm$ SD.
(b) Measured lifetimes of clusters.
The lifetimes for bonded clusters with atom numbers $n = 1$ to $7$ decrease from 10.5(3)~s to 1.8(1)~s.
All data points in this figure are shown as mean values $\pm$ SE.
(c) Predicted lifetimes for larger clusters.
Using the measured one- and three-body loss rates, we estimate the lifetimes of clusters with higher atom numbers. For clusters containing up to 100 atoms, the lifetime is predicted to be approximately 18~ms.
The data are obtained from $784$ samples.
}\label{FigSLifetime}
\end{figure*}

The atomic clusters have sufficiently long lifetimes to allow for the observation of tunneling dynamics.
We measure the lifetime of the Mott insulator states at various filling factors by holding the atoms in deep optical lattices and monitoring the decay of atom number, as shown in Fig.~\ref{FigSLifetime}\rm{a}.
We identify one-body and three-body losses as the primary decay mechanisms, with measured rates of $9.5(3) \times 10^{-2}$ s$^{-1}$ for one-body loss and $2.0(1) \times 10^{-31}$ cm$^6$ s$^{-1}$ for three-body loss (considering the ground-band extension of the atomic Wannier function as a square box).
The one-body loss is attributed to background collisions in the vacuum and light scattering from the optical lattices.
The $1/e$ lifetime of the $n=1$ Mott state is 10.5(3) s, with the short-lattice detuned by only 13 nm from the D2 line contributing to the single-particle loss.


Notably, the observed three-body loss rate is an order of magnitude lower than that reported for strongly interacting one-dimensional gases~\cite{Tolra2004}.
The decay of Mott insulator states at filling factors $n\geq3$ are shown in Fig.\ref{FigSLifetime}\rm{a}, where the initial decay clearly reflects the presence of three-body loss.
The extracted lifetimes are plotted in Fig.\ref{FigSLifetime}\rm{b}, revealing a slower decay as the atom number increases.
For atomic clusters with $n=7$, the measured lifetime exceeds the tunneling period by more than two orders of magnitude (see Fig.\ref{Fig2}\rm{f}).
Based on the measured decay rates, we estimate the expected lifetimes for larger atomic clusters.
As shown in Fig.\ref{FigSLifetime}\rm{c}, even for systems containing up to 100 $^{87}$Rb atoms, the lifetime remains longer than the oscillation periods duration used in Fig.~\ref{Fig2}\rm{f}.

This strong suppression of three-body loss is likely due to the mechanism similar to those in strongly interacting 1D quantum gases~\cite{Tolra2004}, with the on-site interaction strength in our system being even larger, further inhibiting three-body recombination.
The combination of strong tunneling and relatively low decay rates defines a favorable regime for exploring the tunneling dynamics of bonded atoms and opens up opportunities for scaling to even larger cluster sizes and higher total masses.

\section{Atom manipulation and tunneling effects}


\begin{figure*}[htbp]
\includegraphics[width=11.5cm]{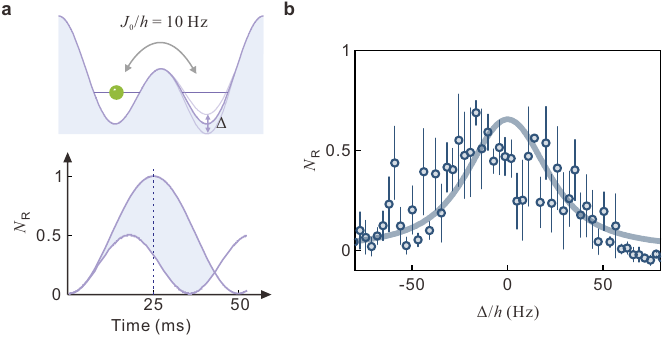}
\caption{Detecting energy fluctuations between left and right wells.
(a) Measurement procedure.
The energy shift $\Delta$ is probed using single-atom tunneling.
With a tunneling strength of $J_0/h = 10.0(2)$~Hz, any small fluctuation in $\Delta$ induces detuning in the Rabi coupling, affecting the tunneling between the two wells.
The amplitude of the atom population is measured at half a tunneling period.
(b) Spectroscopy of single-atom tunneling.
Spectroscopic measurements reveal fluctuations in the atom number on right sites, attributed to noise in $\Delta$.
The spectroscopy is fitted using a Voigt function (solid line), from which the amplitude of the Gaussian noise component is extracted.
The data are obtained from $243$ samples.
All data points in this figure are shown as mean values $\pm$ SEM.
}
\label{FigS3}
\end{figure*}


We conduct quantum tunneling experiment in a balanced double-well superlattice with a phase setting of $\varphi = 0$, as shown in Fig.\ref{Fig2}.
To selectively manipulate atoms in the left or right wells, we introduce a spin-dependent optical superlattice~\cite{Yang2017}.
By tuning the polarization of the short-lattice laser, we apply a vector light shift to the atoms, followed by a microwave pulse to selectively flip the spin state on the right lattice sites, from $\ket{F=1, m_F=-1}$ to $\ket{F=2, m_F=-2}$.
The magnetic bias field is stabilized at 0.82 Gauss with a precision of $\sim$0.1 mG.
A rapid adiabatic passage pulse is employed to flip the spin state, ensuring high-fidelity state preparation.
To initialize the atomic states with only the left sites occupied, we remove atoms from the right sites using a short-time resonant light pulse that only affects the $\ket{F=2, m_F=-2}$ states.

We precisely control and stabilize the superlattice phase to ensure accurate observation of the tunneling effect.
Normally, the lattice depths are set to relatively large values, with $V_{s,l} \sim h \times 50$~kHz, while the energy difference $\Delta$ between the left and right sites should be maintained below the tunneling strength $J_n$, i.e., $\sim h\times 50$ Hz.
This phase stability is crucial for observing tunneling effects of atomic clusters with higher mass.
To minimize phase fluctuations, we stabilize the optical lattice by integrating a significant portion of the retro-reflected optical path within the vacuum chamber, reducing environmental impacts from temperature and humidity variations.

To calibrate the phase stability of the optical lattice, we use single-atom tunneling to measure energy biases between double wells and extract phase fluctuations.
Atoms are initialized to occupy the left sites, with superlattice parameters set to $V_l=8.87(4)  E_r$ and $V_s=27.33(5) E_r$, corresponding to a barrier height of $V_0/k_{\text{B}} = 4.32(1) \ \mu$K and a tunneling strength of $J_0/h = 10.0(2)$ Hz.
After letting the system evolve for 25 ms, equivalent to half a tunneling period, we detect the atoms on the right site.
The resulting spectroscopic measurements, shown in Fig.\ref{FigS3}, exhibit broadening due to phase fluctuations.

The spectroscopy can be well fitted by a Voigt profile, a convolution of a Cauchy-Lorentz distribution and a Gaussian distribution.
The full width at half maximum (FWHM) of the Lorentzian component is 40(1) Hz, determined by the Rabi bandwidth of the tunneling process.
From this, the Gaussian component, arising from fluctuations in the energy bias, is inferred to have an FWHM of 23(1) Hz.
The corresponding relative fluctuation of the energy bias $\Delta/V_0$ is $2.5(1) \times 10^{-4}$, sufficient for tunneling observations in large-mass clusters.

Our \textit{in situ} imaging of the 2D atomic cloud provides insights into the spatial distribution of the superlattice potential. To characterize the spatial variations, we employ an entangled state and perform a Ramsey sequence to quantify energy offsets.
Atoms are initialized in the superposition state $\left( \ket{1,0} + \mathrm{i} \ket{0,1} \right)/\sqrt{2}$ and held for varying durations under lattice depths of $V_s = 30.90(5) E_r$, $V_l = 8.87(4) E_r$, and $V_y = 38.8(3) E_r$. A second Ramsey pulse is then applied, allowing the atoms to tunnel for another quarter period.
The number of atoms that tunnel to the right lattice sites is measured, yielding $N_R$.


\begin{figure*}[htbp]
\includegraphics[width=13.5cm]{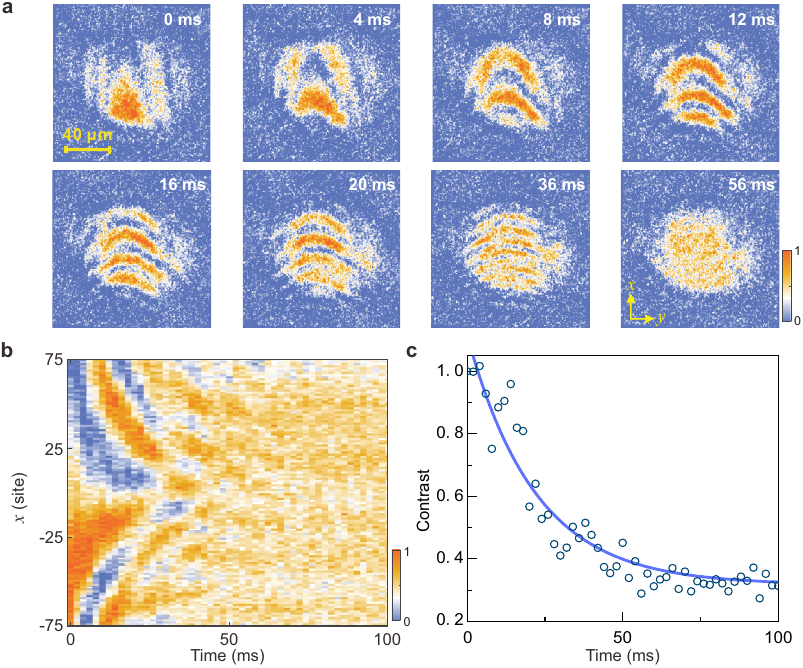}
\caption{Probing the spatial distribution of $\Delta(x,y)$.
We generate single-atom entangled states in double-well units, specifically $\left( \ket{1,0} + \mathrm{i} \ket{0,1} \right)/\sqrt{2}$, and hold them in deep optical lattices for time $t_{\text{R}}$.
In each double-well, the energy bias $\Delta(x,y)$ induces a phase shift to the state, evolving the state to $\left[\ket{1,0} + \exp{\left(\text{i} \left[\Delta(x,y) t_{\text{R}}/\hbar + \pi/2\right]\right)} \ket{0,1} \right]/\sqrt{2}$.
A second $\pi/2$ pulse via tunneling completes the Ramsey sequence for phase measurement.
(a) Spatially resolved Ramsey measurements. The distribution of $\Delta$ in the $x$-$y$ plane is shown at different times, revealing spatial distribution of energy shifts that reflect potential inhomogeneity. Averaging over a certain region leads to effective dephasing of the entangled state.
(b) Time-resolved measurements of the distribution of $\Delta$.
The spatial signal in (A) is averaged over 26 sites along the $y$-direction, revealing inhomogeneity along the $x$-axis. The contrast of the pattern decays over time.
(c) Lifetime of the pattern. To evaluate the contrast of the pattern, we calculate the standard deviation of the measurements from the pattern and plot the contrast as a function of time, revealing a 1/e lifetime of $22(2)$~ms.
}\label{FigS4}
\end{figure*}


This Ramsey method enables high-resolution energy detection. As illustrated in Fig.~\ref{FigS4}, spatial variations in the energy shift $\Delta$ manifest as striped patterns.
In certain double-well units, the phase of the superposition state oscillates over time, indicating an energy imbalance between the two wells.
This imbalance, $\Delta$, exhibits a spatial distribution $\Delta(x,y)$ across the lattice.
By analyzing the standard deviation of atomic density, we estimate a pattern decay lifetime of 22(2) ms, partially attributed to decoherence.
The spatial variation in $\Delta(x,y)$ is around $h\times$45(5) Hz, corresponding to a relative variation of $3.0(3) \times 10^{-4}$ compared to the lattice depth of approximately $h\times$150 kHz along the $x$- or $y$-directions.

Since our measurements are averaged over multiple double-well units, spatial inhomogeneity in the potential limits coherence.
Variations within double wells lead to enhanced apparent decay effects when averaging over larger regions, primarily due to the trapping potential envelope and local disorder.
This explains the reduced coherence time observed when averaging over the entire atomic cloud, as discussed in Ref.~\cite{Folling2007}.
To mitigate spatial inhomogeneity, we employ the 1064 nm pancake lattice to compensate this effect.
Meanwhile, we leverage the spatial resolution of \textit{in situ} imaging to analyze regions with uniform experimental parameters.
This Ramsey method serves as a precise tool for mapping system homogeneity, making our optical lattice platform well-suited for exploring many-body physics.

\begin{figure*}[!htbp]
\includegraphics[width=15.5cm]{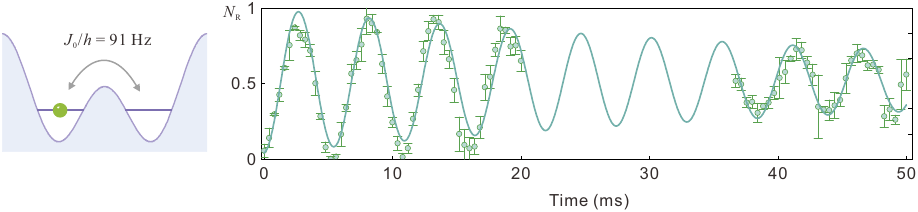}
\caption{Coherence of quantum tunneling.
Coherent tunneling of a single atom is observed at a tunneling rate of $J_0/h=91$~Hz.
The solid curve represents a damped sinusoidal fit, yielding a coherence time of $\tau^{(1)}=58(8)$~ms.
The data are obtained from $307$ samples.
All data points in this figure are shown as mean values $\pm$ SD.
}\label{FigS5}
\end{figure*}

To further understand decoherence mechanisms in the tunneling process, we extract decay times from tunneling dynamics observed in the central region of the cloud.
As shown in Fig.\ref{FigS5}, the oscillatory tunneling behavior is fitted with a damped sinusoidal function to determine the decay time.
Here, an increase in atom number $n$ participating in the tunneling dynamics leads to a reduction in coherence time, as shown in Fig.\ref{Fig2}$\rm{f}$.
The extracted coherence times as a function of atom number are plotted in Fig.\ref{Fig3}\rm{c}.
For a NOON state, which can be viewed as a maximally entangled Greenberger-Horne-Zeilinger (GHZ) state, $\left(\ket{n,0} + \text{i}\ket{0,n}\right)/\sqrt{2}$, any single-particle decoherence leads to the collapse of the entire state.
The energy offset $\Delta$ affects the $n$-th order tunneling process by a factor of $n$, making large clusters more sensitive to energy imbalances.
Fig.\ref{Fig3}\rm{c} shows that the coherence time of tunneling dynamics scales approximately as $1/n$, suggesting that decoherence primarily originates from single-particle effects.
The coherence time of the tunneling oscillation is nearly twice that of the entangled state, as the atoms occupy the entangled configuration only during half of each tunneling cycle.
For clusters of even higher mass, correlated noise sources may introduce additional decoherence, potentially limiting the coherence of such maximally entangled states.

Compared to other high-order processes such as spin-exchange~\cite{Yang2020,Bojovic2025} or ring-exchange dynamics~\cite{Dai2017}, the relatively long coherence times observed in those experiments can be attributed to the absence of atomic center-of-mass motion.
In spin-entangled states, the atomic mass distribution remains balanced across neighboring lattice sites, minimizing sensitivity to spatial energy distributions.
In contrast, the entangled states in our system are spatially separated, with a distance of approximately 320 nm between the left and right wells.
Detecting the NOON state $\left(\ket{n,0} + \text{i}\ket{0,n}\right)/\sqrt{2}$ results in spatially separated outcomes $\ket{n,0}$ or $\ket{0,n}$, whose centers are physically displaced.
This spatial separation increases the susceptibility of the state to energy inhomogeneities, but also enhances its sensitivity to spatial energy variations, making it a valuable resource for precision measurements.

\section{Theoretical simulations}

The seminal work of Ref.~\cite{MacColl1932} established that the transmission probability of particles through a barrier decreases exponentially with mass, following the relation of $\exp(-\gamma \sqrt{m})$.
In our optical lattice system, the tunneling strength for single atoms or tightly bound molecules can be calculated using band theory.
In the tight-binding regime, the tunneling strength $J_0$, equivalent to a quarter of the ground-band bandwidth, is given by,
\begin{equation}
\begin{split}
J_0 &\simeq \frac{4}{\sqrt{\pi}} E_r \left( \frac{V_0}{E_r} \right)^{3/4}
\exp \left[ -2 \left( \frac{V_0}{E_r} \right)^{1/2} \right].
\end{split}
\end{equation}\label{eq:J0}
This leads to the same exponential dependence of the tunneling strength on both the particle mass and the barrier height,
\begin{equation}
J_0 \propto (m)^{-1/4} \exp \left[ -\gamma \sqrt{m} \right],~\gamma = \frac{2 \sqrt{2 V_0} \lambda_s}{h}.
\end{equation}\label{eq:J0prop}
This exponential decay of tunneling strength, even in ultracold atomic systems, presents significant challenges for observing quantum tunneling of objects with large masses.
Prior to this work, this behavior was generally considered an intrinsic and invariable property.

We propose that when atoms form weakly bonded clusters, the tunneling strength transitions from an exponentially suppressed regime to near-unity scaling.
This transition can be understood using high-order perturbation theory.
In the regime where the tunneling strength $J_0$ is much smaller than the on-site interaction $U$ ($J_0 \ll U$), the effective Hamiltonian for the clusters can be derived. Taking a seven-atom cluster as an example, the tunneling strength is determined by the matrix element,
\begin{equation}
    \bra{0,7} \hat{H}_{\text{c}} \ket{7,0} = \frac{\prod_{i=0}^{6} \langle 6-i,i+1|  \hat{H_0}| 7-i,i \rangle}{\prod_{i=1}^{6} \left(E_{| 7,0\rangle}-E_{| 7-i,i\rangle}\right)}.
\end{equation}\label{eq:Pertubation}
As shown in Fig.\ref{Fig2}\rm{c}, 6 intermediate virtual states couple the initial state $\ket{7,0}$ to the final state $\ket{0,7}$.
The prefactor $\alpha_n$ decreases with increasing cluster size due to the growing energy detuning of intermediate virtual states, where the energy gap for an $n$-atom Mott insulator is $\left(n-1\right)U$.
Meanwhile, bosonic enhancement increases the coupling strength by a factor of $n!$.
These competing effects yield an overall coefficient of $n/(n-1)!$ for an $n$-atom cluster.
In contrast to strongly bound molecular clusters, whose tunneling is suppressed by strong localization, weakly bound clusters can tunnel collectively through high-order perturbative processes.
{The coupling strength of this high-order tunneling process can be derived as follows.}

\begin{equation}
{
\begin{aligned}
 \bra{0,n} & \hat{H}_{\text{c}} \ket{n,0} =\frac{\prod_{i=0}^{n-1}\langle n-1-i,\; i+1 \vert \hat H_0 \vert n-i,\; i \rangle}{\prod_{i=1}^{n-1}\left( E_{n,0} - E_{n-i,i} \right)}\\
=&
\frac{\prod_{i=0}^{n-1}\left( \sqrt{n-i}\,\sqrt{i+1}\, J_0 \right)}{\prod_{i=1}^{n-1}\left[n(n-1)-(n-i)(n-i-1)-i(i-1)\right]U/2}\\
=&\frac{n}{(n-1)!}\frac{J_0^n}{U^{n-1}}
\end{aligned}}
\end{equation}


\begin{figure*}[htbp]
\includegraphics[width=13.8cm]{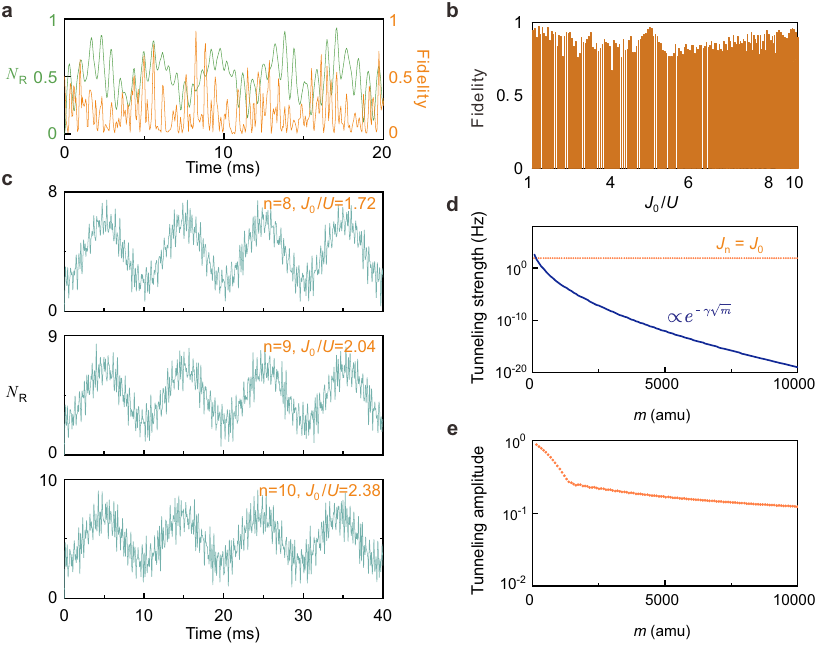}
\caption{Theoretical simulations.
(a) Tunneling dynamics of the seven-atom cluster.
We simulate the tunneling dynamics at a coupling parameter of $J_0/U = 1.92$.
The $N_{\text{R}}$ is plotted as the green curve, while the orange curve shows the fidelity for preparing the entangled state $\left( \ket{7,0} + \mathrm{i} \ket{0,7} \right)/\sqrt{2}$.
(b) Fidelity of NOON states at various coupling strengths.
As the coupling strength $J_0/U$ increases, the system moves away from the perturbation regime.
Despite this, the fidelity for realizing the NOON state remains considerably high even at $J_0/U = 10$.
(c) Tunneling dynamics for larger clusters ($n=8, 9, 10$).
For these clusters, we set $J_0/U$ at proper values to achieve tunneling oscillations with appropriate frequencies.
(d) Tunneling strength for larger clusters.
Using high-order perturbation theory, the tunneling strength is given by $\alpha_n J_0^n / U^{n-1}$.
Assuming the cluster mass increases up to $m = 10^4$~amu, one can maintain comparable tunneling strengths across different atom numbers by appropriately tuning the coupling ratio $J_0/U$.
In contrast, the tunneling strength for strongly bound molecular clusters decreases sharply, following the conventional exponential dependence on mass, as illustrated by the blue curve.
(e) Tunneling amplitude.
For the atomic clusters shown in Fig.\ref{FigS6}\rm{d}, the corresponding tunneling amplitude remains above 0.1, even for the largest masses considered.
}\label{FigS6}
\end{figure*}


For weakly bonded clusters, tunneling dynamics can be accurately computed using exact diagonalization of the Bose-Hubbard Hamiltonian.
Taking the seven-atom cluster as an example,
the Hamiltonian matrix $\hat{H}_0 $ is given by,
\begin{equation}
\setlength{\arraycolsep}{0.1pt}
\renewcommand{\arraystretch}{0.9}
\left(
\scriptsize
\begin{array}{cccccccc}
21U & -\sqrt{7}J_0 & 0 & 0 & 0 & 0 & 0 & 0 \\
-\sqrt{7}J_0 & 15U & -\sqrt{12}J_0 & 0 & 0 & 0 & 0 & 0 \\
0 & -\sqrt{12}J_0 & 11U & -\sqrt{15}J_0 & 0 & 0 & 0 & 0 \\
0 & 0 & -\sqrt{15}J_0 & 9U & -4J_0 & 0 & 0 & 0 \\
0 & 0 & 0 & -4J_0 & 9U & -\sqrt{15}J_0 & 0 & 0 \\
0 & 0 & 0 & 0 & -\sqrt{15}J_0 & 11U & -\sqrt{12}J_0 & 0 \\
0 & 0 & 0 & 0 & 0 & -\sqrt{12}J_0 & 15U & -\sqrt{7}J_0 \\
0 & 0 & 0 & 0 & 0 & 0 & -\sqrt{7}J_0 & 21U
\end{array}
\right)
\label{eq:H0matrix}
\end{equation}

We numerically evolve the system under the Hamiltonian $\ket{\psi(t)} = \exp{\left(-\text{i}\hat{H_0}t/\hbar\right)} \ket{\psi(0)}$ in the Fock-state basis.
For multi-atom tunneling, density-induced tunneling effects further enhance the tunneling strength as the atom number increases.
We describe the system using a density matrix $\rho$, and quantify the fidelity of the NOON state as
$\mathcal{F}=\langle \psi_{\text{NOON}}|\rho_\text{exp}|\psi_{\text{NOON}}\rangle$, where for a seven-atom cluster $\ket{\psi_{\text{NOON}}} = \left( \ket{7,0} + \mathrm{i} \ket{0,7} \right)/\sqrt{2}$.

In the absence of decoherence, Fig.\ref{FigS6}\rm{a} shows the fidelity dynamics of the seven-atom NOON state.
Multiple frequency components contribute to the oscillations, and constructive interference at specific times leads to local maxima in fidelity.
Fig.~\ref{FigS6}\rm{b} presents the maximum achievable fidelity across different $J_0/U$ ratios.
Notably, a fidelity of 86.4\% can be attained even at $J_0/U = 3$.
In the presence of decoherence, however, the optimal point in the oscillation curve must be reached within the coherence time window.


\begin{figure*}[!htbp]
\includegraphics[width=13.2 cm]{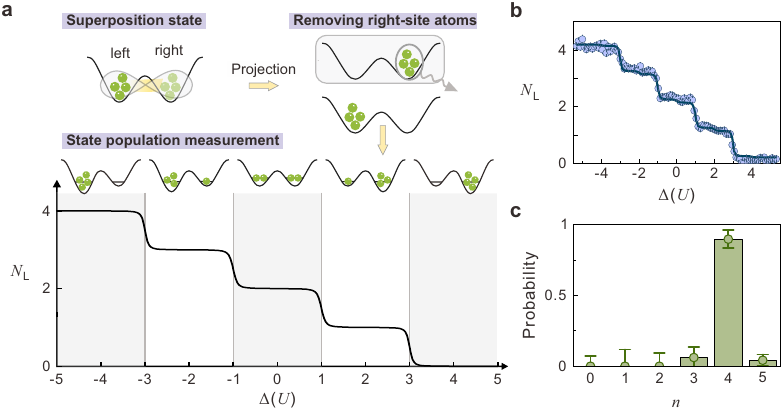}
\caption{{
Full counting statistics of site occupation.
(a) Measurement procedure. For the atomic clusters in the superposition states, the projection results in Fock state in left and right wells. By removing the right-site atoms, we probe the probabilities of site occupation of the left well. As the energy shift $\Delta$ is tuned, the atom will form atomic configuration sketched above the curve. Using the state $\ket{4,0}$ as an example, the lower panel shows the number of atoms on the left sites as a function of $\Delta$.
(b) Counting the number distribution. For the initial state of 4-atom clusters, we measured atom number on the left site which shows a step-like behavior of the site occupation by tuning the $\Delta$.
The solid line is the corresponding fitting results.
Error bars represent the $\pm 1 \sigma$ statistical standard deviation of the measurements.
(c) From the fitting curve in (b), we directly extract the probabilities for each site occupations, giving the full counting statistical results.
The data are obtained from $303$ samples.
All data points in this figure are shown as mean values $\pm$ SEM.
}
}\label{FigS8}
\end{figure*}


The experimental results are well captured by the theoretical model described above.
As shown in Fig.\ref{Fig2}\rm{f}, the primary low-frequency oscillations correspond to the tunneling dynamics of $n$-atom bonded clusters.
Superimposed high-frequency components arise from residual lower-order tunneling processes.
Imperfect preparation of the Mott insulator leads to additional occupancies in some double wells.
These are accounted for by summing the contributions of all possible configurations, weighted by their measured population fractions.
Slight deviations in the prefactor $\alpha_n$, as shown in Fig.\ref{Fig3}\rm{a}, arise when the parameters move beyond the perturbative regime, where $J_0/U$ approaches unity and the energy gap between the eigenstates is shifted.
Additionally, the coherence times extracted from damped sinusoidal fits are included in the theoretical model by applying a decay rate to the off-diagonal elements of the density matrix.

To explore tunneling dynamics in larger clusters, we extend our simulations to higher masses (e.g., $n = 8, 9, 10$).
Assuming an on-site interaction of $U/h = 800$~Hz, we adjust the coupling parameters $J_0/U$ to optimize the tunneling strength.
As shown in Fig.~\ref{FigS6}\rm{c}, the dominant oscillations correspond to the tunneling of bonded clusters.
Despite the presence of higher-frequency oscillations arising from lower-order processes, the collective tunneling of the entire cluster remains within the experimentally accessible regime.
NOON states maintain high fidelity around the quarter-period evolution, confirming the feasibility of scaling to 10 atoms with a total mass of 869 amu.

Assuming a sufficiently large tuning range for the Hubbard parameters $J_0$ and $U$, we calculate the tunneling strength and oscillation amplitude for much larger atomic clusters.
By setting the effective tunneling strength $J_n = J_0$, we find that the coupling ratio satisfy $J_0/U = \left[(n-1)!/n\right]^{1/(n-1)}$.
In Fig.~\ref{FigS6}\rm{d}, we compare the results of this calculation with the tunneling predictions for strongly bonded molecules.
In parallel, we compute the oscillation amplitude directly using the Hubbard model by evaluating the wavefunction overlap between the eigenstates and the target NOON state.
As shown in Fig.~\ref{FigS6}\rm{e}, for clusters with total mass up to $10^4$ amu, the normalized oscillation amplitude remains above 0.1.
This decay is significantly slower than the typical exponential suppression with mass, indicating the feasibility of coherent tunneling in such large systems.

It is well known that extending to larger clusters introduces additional constraints: the entire on-site repulsion may exceed the barrier height, invalidating the two-body interaction approximation, increasing three-body loss rates, and reducing coherence time.
Nevertheless, our analysis reveals a parameter regime where the tunneling of high-mass clusters remains observable, provided that the coherence time can be extended to levels demonstrated in recent cavity-enhanced lattice systems~\cite{Panda2024}.

{\section{Collective tunneling and full counting statistics}}

\begin{figure*}[htbp]
\includegraphics[width=11.9 cm]{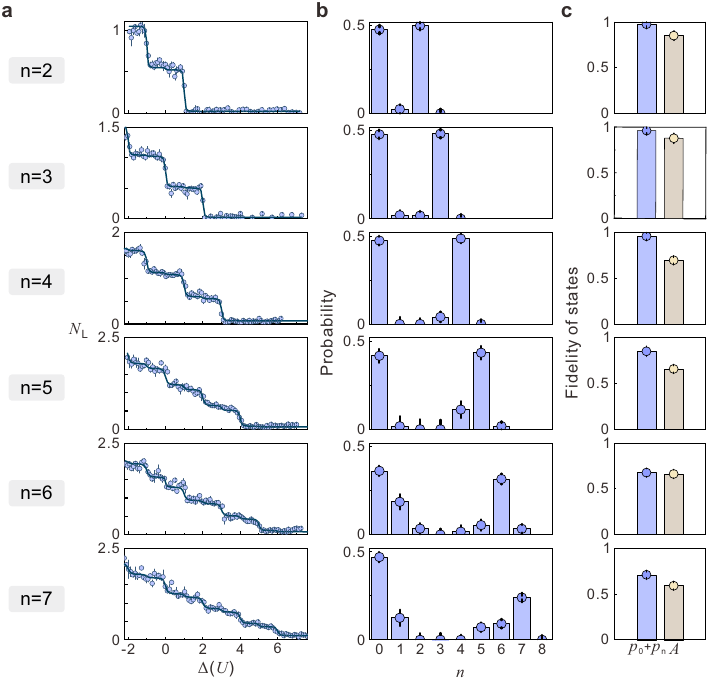}
\caption{{
Counting statistics for the superposition states.
(a) Measurement of left-site number probabilities.
For superposition states prepared after a quarter tunneling period with atom numbers ranging from $n = 2$ to $n=7$, we project the states onto the Fock basis and measure the left-site number distribution as a function of the energy offset $\Delta$.
Error bars represent the $\pm 1 \sigma$ statistical standard deviation of the measurements.
(b) Atom-number probabilities.
The atom-number probabilities are extracted from the data in (a) by fitting the step-like curves, with $p_n$ as fitting parameters. Occupation components from 0 to $n+1$ atoms are resolved.
(c) State fidelity.
The probabilities $p_0$ and $p_n$ correspond to the two components contributing to the fidelity of the NOON state. We plot their sum together with the measured tunneling oscillation amplitude $A$ shown in Fig.~\ref{Fig3}\rm{c}. The summed occupation probabilities are generally larger than the oscillation amplitudes.
The data are obtained from $1544$ samples.
All data points in this figure are shown as mean values $\pm$ SEM.
}
}\label{FigS9}
\end{figure*}


{Our experiment is not operated deep in the perturbative regime where $J_0 \ll U$, and therefore a finite leakage into unwanted states beyond $\ket{n,0}$ and $\ket{0,n}$ is in principle possible.
To characterize this effect, we measure both the mean atom number and the number fluctuations (see Fig.~\ref{FigSMI}), which confirm the intended number distribution of the initial states.
During the tunneling dynamics, although theoretical calculations indicate that the leakage remains small and can be controlled by choosing an appropriate tunneling duration, it is nevertheless essential to directly measure the actual population distribution.
Evaluating the state fidelity from the site-occupation probabilities provides an independent and direct confirmation of the indivisible tunneling of atomic clusters.}

{To this end, we perform measurements of the exact site occupation of the superposition states and extract the probabilities $p_n$ for each occupation number.
After a quarter tunneling period, the atomic clusters form a quantum superposition state.
To probe the population distribution, we first project the state by ramping up the double-well barrier to freeze the dynamics, thereby mapping the superposition onto Fock states in the left and right wells.
As shown in Fig.~\ref{FigS8}\rm{a}, we then remove the atoms on the right sites and measure the site occupation on the left well.
Using full counting statistics of the left well and invoking atom-number conservation, we can directly reconstruct the atom-number distribution of the superposition states.}

{To resolve the site-occupation probabilities from $0$ to $n+1$, we employ a full counting statistics method based on the nonlinear on-site interaction, which renders the on-site energy a unique fingerprint of the atom number~\cite{Cheinet2008,Kwan2024}.
In the double-well superlattice, we control the energy offset $\Delta$ between the two wells to discriminate different occupation numbers, as illustrated in Fig.~\ref{FigS8}\rm{a}.
When $\Delta$ matches the energy gap between the states $\ket{n-i,i}$ and $\ket{n-i-1,i+1}$ (with integer $i$ ), resonant tunneling occurs and the system reconfigures.
Taking $n=4$ as an example, Fig.~\ref{FigS8}\rm{a} shows five distinct plateaus as $\Delta$ is varied from  $-4U$ to $4U$,  corresponding to the $n+1=5$ possible configurations.
For each atom number $n$, a characteristic sequence of plateaus emerges as a function of $\Delta$.}

{We validate this full counting statistics method by first measuring a known reference state, the initial state $\ket{4,0}$.
In an adiabatic protocol, atoms are transferred into the long-lattice site by slowly reducing the short-lattice depth $V_s$ to 0, and subsequently ramping it up to $V_s = 44 E_r$ over 20 ms.
The superlattice phase $\varphi$ controls the energy offset $\Delta$, which is fixed for each experimental realization.
We then measure the remaining atoms on the left site as a function of $\Delta$ as shown in Fig.~\ref{FigS8}\rm{b}.
Five well-resolved plateaus are clearly visible, indicating that a large fraction of clusters occupy the $\ket{4,0}$ state.
Fitting the data yields the occupation probabilities shown in Fig.~\ref{FigS8}\rm{c}, where small contributions at $n\pm1$ are observed, consistent with the number-fluctuation measurements presented in Fig.~\ref{Fig2}\rm{d} of the main text.}

{We then apply the same full counting statistics technique to the superposition states for atom numbers ranging from $n=2$ to $n=7$.
After waiting for a quarter tunneling period, we project the state and measure the atom-number probabilities.
As shown in Fig.~\ref{FigS9}\rm{a}, the characteristic step-like dependence on $\Delta$ is clearly resolved for each atomic cluster.
Owing to the symmetry of the configurations for positive and negative $\Delta$ we primarily analyze the data for positive offsets.
As the cluster size increases, finite-temperature effects lead to enhanced particle and hole excitations, manifesting as populations in the $n \pm 1$ states.
Because projection of the superposition state leaves approximately half of the left sites empty, the probability $p_0$ is comparable to $p_n$, as shown in Fig.~\ref{FigS9}\rm{b}.
Importantly, leakage into unwanted states beyond $\ket{n,0}$ and $\ket{0,n}$ remains consistently small, with the dominant leakage occurring in the nearest neighboring states $\ket{n-1,1}$ and $\ket{1,n-1}$.}

{From these measurements, we conclude that the population leakage contributes to the reduction of the oscillation amplitude observed in the Ramsey interferometry data.
By comparing the total population $p_0 + p_n$ with the measured oscillation contrast, we find that the occupation probability is systematically larger, indicating that dephasing is another dominant mechanism responsible for the decay of the tunneling oscillations.}

\vspace*{0.5cm}

\section{Spatial superposition states and quantum enhanced measurements}

\begin{figure*}[htbp]
\includegraphics[width=15.5 cm]{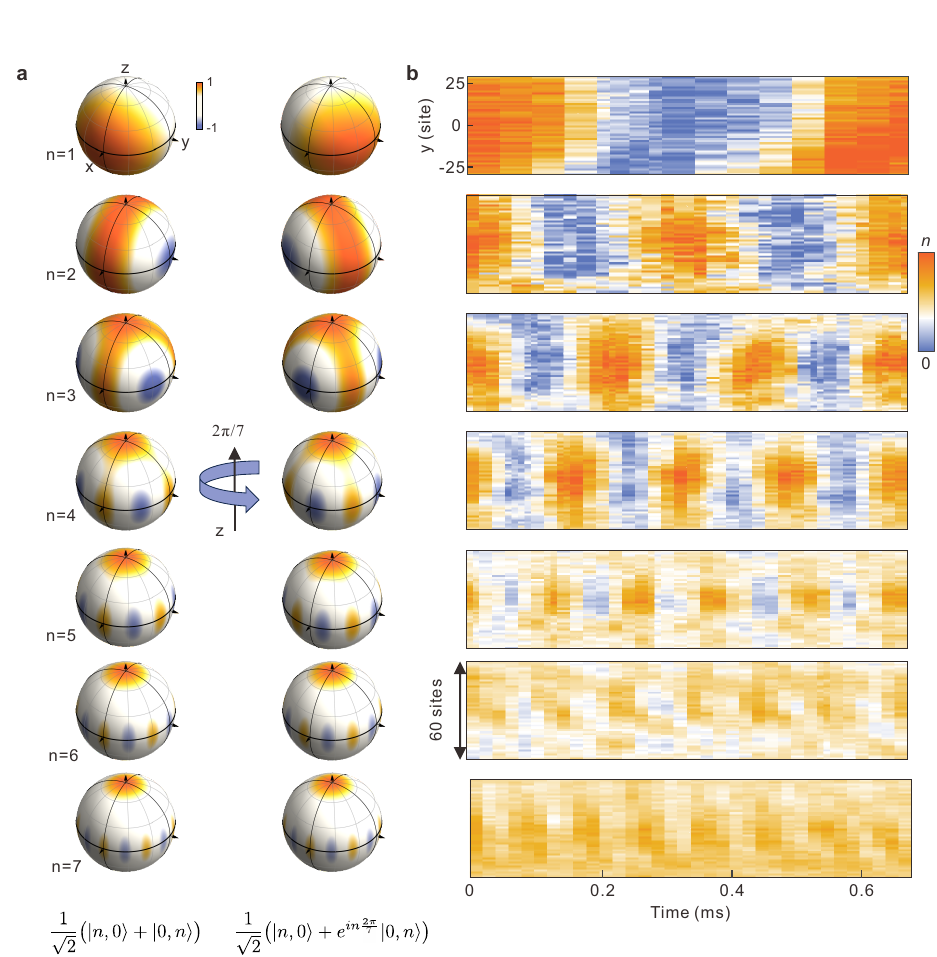}
\caption{Spatial distribution of Ramsey interferometry.
(a) The normalized Wigner distribution on the Bloch sphere illustrates the ideal NOON state for various values of $n$.
The distribution demonstrates that rotating the state by $2\pi/n$ around the $z$-axis restores the initial $n$-atom NOON state.
The left panel shows the initial NOON states, while the right panel displays their distributions after a $2\pi/7$ rotation.
(b) Ramsey interferometry.
Phase information is extracted from the atomic density on the right sites after the Ramsey sequence, revealing clear spatial patterns and oscillations. This enables detection of micrometer-scale energy shifts.
}\label{FigS10}
\end{figure*}


Precision measurements can reach their ultimate limit through quantum entangled states, such as the NOON state, which exploits quantum superpositions of $N$ indistinguishable particles.
This enables measurements to approach the Heisenberg limit, achieving precision $\Delta \theta = 1/N$ that surpasses the standard quantum limit of $\Delta \theta = 1/\sqrt{N}$.
NOON states have been successfully realized experimentally in photon~\cite{Walther2004, Mitchell2004,Nagata2007,Afek2010,Thomas2022} and phonon~\cite{Zhang2018} systems.
However, prior to this work, the creation of massive NOON states involving more than two atoms has yet to be achieved.
Unlike photons, massive particles are more susceptible to a wider range of interactions, making the generation of NOON states—especially spatially separated ones—considerably more challenging.
However, this sensitivity also enables significant enhancements in measurement precision, with potential applications in gravity sensing~\cite{Pezze2018} and fundamental tests of quantum mechanics~\cite{Bassi2013}.

We generate the entangled state through quantum tunneling, with the fidelity inferred from its oscillation amplitude.
The atomic clusters are initially prepared on the left sites and allowed to tunnel for a quarter period.
During tunneling, the atoms bond into clusters and tunnel as a single entity, avoiding distinct population of intermediate virtual states.
Even in the $J_0/U \sim 1$ regime, high-fidelity NOON states with large $n$ can be achieved by optimizing the tunneling time, see Fig.\ref{FigS6}\rm{b}.

In the experiment, population leakage to unwanted states and decoherence effects reduce the fidelity of  the target superposition state $\left(\ket{7,0} + \mathrm{i} \ket{0,7}\right)/\sqrt{2}$.
Taking into account imperfections in the Mott insulator and fluctuations in experimental parameters, theoretical simulations that incorporate these effects yield a fidelity consistent with the Ramsey measurements. The oscillation amplitude observed in Fig.\ref{Fig2}\rm{f} is 0.59(4), consistent with the Ramsey measurement of 0.36(2) shown in Fig.\ref{Fig4}\rm{b} (two times of the $\pi/2$ operations).

To characterize the states for quantum metrology, we utilize Fisher information to assess its precision (Fig.~\ref{Fig4}).
Fisher information is generally expressed as~\cite{Pezze2018},
\begin{equation}
F(\theta)=\sum_{\mu} \frac{1}{P(\mu |\theta)}\left( \frac{\partial P(\mu |\theta)}{\partial \theta}\right)^2,
\end{equation}\label{eq:Fisher}
where $\mu$ denotes the eigenvalue of an observable and $P(\mu |\theta)$ represents the associated probability.
In our system, the parity of the atoms serves as the observable, which means right site has the parity +1, and the left site has parity -1.
For a pure NOON state, $\ket{\psi_\text{NOON}}=\left[\ket{n,0} + e^{\text{i}(n\theta +\pi/2)}\ket{0,n}\right]/\sqrt{2}$, the probability after applying the second Ramsey $\pi/2$ pulse is,
\begin{equation}
P(\mu = \pm 1|\theta)=\frac{1\pm \cos n \theta}{2}.
\end{equation}\label{eq:Parity1}

The Fisher information for this pure NOON state is $F(\theta)=n^2$, which is $n$ times the standard quantum limit.
In this context, the quantum Cram\'{e}r-Rao bound~\cite{Braunstein1994,Pezze2018} gives the maximum phase sensitivity as, $\Delta\theta = 1/\sqrt{F(\theta)}=1/n$.
In practice, experimental noise and decoherence reduce the Fisher information, which is related to the visibility of the parity oscillation.
In our Ramsey measurements, the parity oscillation amplitude is,
\begin{equation}
P(\mu = \pm 1 |\theta)=\frac{1\pm A \cos n \theta}{2},
\end{equation}\label{eq:Parity2}
where $0 \le A \le 1$ is the visibility.
The Fisher information is then given by,
\begin{equation}
F(\theta)= \frac{A^2 n^2 \sin^2n\theta}{1-A^2 \cos^2n\theta}.
\end{equation}\label{eq:FisherTheta}
The Fisher information reaches its maximum value $F = A^2 n^2$ when $\sin n\theta = \pm 1$.

We demonstrate quantum precision measurements beating the standard quantum limit using Fisher information.
A spatially distributed staggered potential is introduced via a spin-dependent optical superlattice, inducing an energy bias of $\Delta/h =1.49(2)$~kHz between the left and right sites.
The entangled states evolve in this staggered potentials, accumulating a phase $n\theta$.
The phase is then projected onto the atom number imbalance between the two sites through a second Ramsey $\pi/2$-pulse.

From the measurements in Fig.~\ref{Fig4}\rm{b}, we extract the oscillation amplitude of the parity (imbalance) $A$ and derive the Fisher information $F = A^2 n^2$.
The standard quantum limit corresponds to measurements with $n$ copies of the superposition state $(\ket{1,0} + \mathrm{i}\ket{0,1})/\sqrt{2}$.
Using single entangled NOON states, the Fisher information values and the achieved derived phase sensitivities are listed for different atom number $n$:
\begin{align}
n=1:&\; F=0.87(6),\ \Delta\theta=1.07(3); \nonumber \\
n=2:&\; F=2.5(1),\ \Delta\theta=0.63(2); \nonumber \\
n=3:&\; F=5.7(2),\ \Delta\theta=0.42(1); \nonumber \\
n=4:&\; F=8.8(4),\ \Delta\theta=0.34(1); \nonumber \\
n=5:&\; F=7.9(9),\ \Delta\theta=0.36(2); \nonumber \\
n=6:&\; F=5.9(6),\ \Delta\theta=0.41(2); \nonumber \\
n=7:&\; F=2.6(4),\ \Delta\theta=0.61(4). \nonumber
\end{align}

For the four-body NOON state, this results in a 1.7(1)~dB enhancement in phase sensitivity beyond the standard quantum limit, confirming the improved precision achievable in quantum metrology.


It is important to note that the sub-Poissonian number distribution of our initial states serves as a key quantum resource.
In contrast, coherent states with Poissonian statistics are limited to precision at the standard quantum limit.
The generation of NOON states in our experiment relies critically on the sub-Poissonian atom number statistics realized via atomic Mott insulators.
With sufficiently large array of atomic clusters, these results demonstrate that our cold-atom platform offers a scalable and practical route for quantum sensing of spatially varying potentials.

Finally, we observe spatially distributed oscillation patterns in quantum precision measurements (Fig.~\ref{FigS10}\rm{b}).
The central region shows the expected frequency scaling linearly with $n$, while phase shifts occur at the periphery, due to preparation inhomogeneities and spatial variations in the energy field.
The oscillation frequency can be determined with high accuracy, making the NOON state suitable for measuring various interactions and energy shifts.
To further enhance precision, optical tweezers could be employed to separate the atomic cluster into larger distance~\cite{Kaufman2014,Murmann2015}, thereby improving sensitivity to fields such as gravity.

\end{document}